\documentclass[10pt,preprint]{aastex}

\usepackage{lscape}

\begin{document}

\title{Lost and found: evidence of second generation stars along the
asymptotic giant branch of the globular cluster
NGC~6752}\footnotetext[1]{Based on observations collected at the
ESO-VLT under the program 095.D-0320(A).}

\author{
E. Lapenna\altaffilmark{1,2},
C. Lardo\altaffilmark{3},
A. Mucciarelli\altaffilmark{1},
M. Salaris\altaffilmark{3},
F. R. Ferraro\altaffilmark{1},
B. Lanzoni\altaffilmark{1},
D. Massari\altaffilmark{2,4},
P. B. Stetson\altaffilmark{5},\\
S. Cassisi\altaffilmark{6},
A. Savino\altaffilmark{3,4}
}

\affil{\altaffilmark{1} Dipartimento di Fisica e Astronomia, Alma Mater Studiorum Universit\`a di Bologna,
Viale Berti Pichat 6/2, I--40127 Bologna, Italy \\
\altaffilmark{2} INAF - Osservatorio Astronomico di Bologna, Via Ranzani 1 - 40127 Bologna, Italy\\
\altaffilmark{3} Astrophysics Research Institute, Liverpool John Moores University, 146 Brownlow Hill, Liverpool L3 5RF, United Kingdom \\
\altaffilmark{4} Kepteyn Astronomical Institute, University of Groningen, Landleven 12, 9747 AD Groningen, The Nethelands \\
\altaffilmark{5} National Research Council, 5071 West Saanich Road, Victoria, BC V9E 2E7, Canada \\
\altaffilmark{6} INAF - Osservatorio Astronomico di Teramo, via Mentore Maggini, 64100, Teramo, Italy \\
}
%

%\date{}

\begin{abstract}
  We derived chemical abundances for C, N, O, Na, Mg and Al in 20
  asymptotic giant branch (AGB) stars in the globular cluster NGC
  6752. All these elements (but Mg) show intrinsic star-to-star
  variations and statistically significant correlations or
  anticorrelations analogous to those commonly observed in red giant stars 
  of globular clusters hosting multiple populations.
  This demonstrates that, at odds with previous findings, both first and second generation stars
  populate the AGB of NGC 6752.  The comparison with the Na abundances
  of red giant branch stars in the same cluster reveals that
  second generation stars (with mild Na and He enrichment) do
  reach the AGB phase.  The only objects that are not observed along
  the AGB of NGC 6752 are stars with extreme Na enhancement.  This is
  also consistent with standard stellar evolution models, showing that
  highly Na and He enriched stars populate the bluest portion of the
  horizontal branch and, because of their low stellar masses, evolve
  directly to the white dwarf cooling sequence, skipping the AGB
  phase.
\end{abstract}


%%%%%%%%%%%%%%%%%%%%%%%%%%%%%%%%%%%%%%%%%%%%%%%%%%%%%%%%%%%%%%%%%%%%%%%% INTRO

\section{Introduction}

The vast majority of Galactic globular clusters (GCs) host multiple
stellar populations (MPs) characterized by different abundance ratios
of selected light elements \citep[see, e.g.,][for a review]{gratton12}
: some stars share the same light element abundance ratios
measured in Galactic field stars with similar metallicity, but a
large fraction of the cluster population has enhanced N, Na, and Al
and depleted C and O abundances.
The patterns are not random, but anticorrelated variations of the
pairs C-N and O-Na are commonly observed. These are generally
considered to arise from hot hydrogen burning in a previous generation
of more massive stars, as asymptotic giant branch (AGB) stars
\citep{ventura05}, fast-rotating massive stars \citep{decressin07},
interacting massive binary stars \citep{demink09}, and/or
super-massive stars \citep{denis14}.\footnote{We refer
the reader to \cite{renzini15} for a critical analysis of the
various scenarios for the polluters.}  Objects with standard
composition are commonly denoted as first generation (FG) stars, and
those with modified chemistry as second generation (SG) stars,
although the assumption that they are formed in subsequent star
formation episodes is sometimes questioned \citep[see, e.g.,][]{bastian13}.

In a few GCs the SG/FG star ratio measured along the red giant branch
(RGB) is observed to differ from that measured along the AGB, with a
substantial deficiency of SG stars within the AGB population, compared
to the RGB \citep{norris81, gratton10a, campbell12,campbell13,
  johnson15,lapenna15,maclean16}. In principle, this can be
  explained by taking into account that stars with evolving masses
  below $0.55 M_\odot$ are expected to fail reaching the AGB phase
  (the so-called {\sl AGB-manqu\'e} stars; see, e.g., \citealp{gr})
  and SG stars are indeed expected to have a lower mass along the HB with
respect to FG stars.  In fact, since they are typically He-enhanced,
they originate from RGB stars with a lower mass and end up, on
average, with a lower mass along the HB, if the RGB mass loss is
approximately the same for FG and SG sub-populations \citep[see
e.g.][]{cs}.  One therefore expects that the AGB of GCs with a blue
HB should lack at least part of the SG component, compared to what is
seen along the RGB. This is consistent with the findings of
\citet{gratton10a}, who empirically showed that the number ratio
between AGB and HB stars (the $R_2$ parameter) correlates
with the HB morphology, in the sense that clusters with the bluest HB
morphology have lower $R_2$ values.

NGC 6752 is a metal-intermediate GC with an extreme blue HB morphology
and a low $R_2$ value, and it is claimed to be the most extreme case
of a GC lacking SG stars along the AGB.
In fact, \citet[][hereafter C13]{campbell13} measured the Na abundance
of 20 AGB stars in this cluster and from the derived [Na/Fe]
distribution,
they concluded that all objects belong to the FG population.
  In their interpretation, the SG stars fail to reach the AGB phase
  because their HB progenitors are all located at effective
  temperatures ($T_\mathrm{eff}$) hotter than the Grundahl Jump (at
  $\sim$11 500 K) and experience a very strong mass loss (a factor of
  20 larger than that suffered along the RGB).\footnote{We note that
  this  assumption is at odds with the constraints from the stellar wind
  models for HB stars provided by \cite{vink02}.} An alternative
  solution has been proposed by \citet{charbonnel13}, who argued that
  the lack of SG AGB stars can be explained within the fast-rotating
  massive stars scenario by assuming very high He abundances (up to
  $Y\sim$0.7) for the SG objects, that therefore become {\sl
    AGB-manqu\'e} stars.  On the other hand, by using detailed
  synthetic HB simulations, \citet{cassisi14} were able to reproduce
  the star distribution along the HB of NGC 6752 and its observed
  $R_2$ value assuming the initial He-abundance distribution derived
  from the cluster main sequence \citep[$Y$ between $\sim$0.25 and$\sim$0.27; see][]{milone13}
  without invoking any extreme HB mass loss or initial He
  enhancement. However, these simulations show that $\sim 50\%$ of the
  AGB population should be composed of SG stars, at odds with the
  claim by C13.

  With the aim of solving this intriguing issue, here we present
  the chemical abundances of iron and several light elements that we
  recently determined from high-resolution spectra for the same sample
  of AGB stars discussed in C13.

%%%%%%%%%%%%%%%%%%%%%%%%%%%%%%%%%%%%%%%%%%%%%%%%%%%%%%%%%%%%%%%%%%%%%%%% OBSERVATIONS

\section{Observations}
The 20 AGB stars in NGC 6752 previously studied by C13 have been
re-observed (program 095.D-0320(A), PI: Mucciarelli) with the
UVES spectrograph \citep{dekker00} mounted at the ESO-Very Large
Telescope. We used the Dichroic1 mode adopting the gratings 390 Blue
Arm CD\#2 and 580 Red Arm CD\#3 with the 1 arcsec slit (R=~40000).
Exposure times range from $\sim$10 min for the brightest targets to
$\sim$25 min for the faintest ones, to obtain pixel signal-to-noise
ratios higher than 100.  The data reduction was performed by using the
dedicated ESO pipeline, including bias subtraction, flat fielding,
wavelength calibration, spectral extraction and order merging.

%%%%%%%%%%%%%%%%%%%%%%%%%%%%%%%%%%%%%%%%%%%%%%%%%%%%%%%%%%%%%%%%%%%%%%%% CHEMICAL ANALYSIS

\section{Chemical analysis}
\label{ca}

The chemical analysis has been performed following the same procedure described in \cite{lapenna15}.
The stellar atmospheric parameters have been derived as follows:\\ 
(1)~$T_\mathrm{eff}$ have been derived spectroscopically by requiring no trend between iron abundances and 
excitation potentials;\\ 
(2)~surface gravities (log$~g$) have been obtained through the Stefan-Boltzmann relation,  
adopting the spectroscopic $T_\mathrm{eff}$, the distance modulus ($m-M$)$_{0}$ = 13.13 
and color excess E($B-V$) = 0.04 \citep{ferraro99}, and a mass of 0.61 M$_{\odot}$,
according to the median value of the HB mass range estimated by 
\citet{gratton10b}\footnote{\citet{cassisi14} 
derived a slightly lower ($\sim$0.55 M$_{\odot}$) median mass. 
The adoption of this value decreases log~g by $\sim$0.04, with a negligible impact on the abundances, 
$\sim$0.02 for [FeII/H] and smaller than 0.01 dex for the other species.}. 
Stellar luminosities have been calculated using  the bolometric corrections by \citet{alonso99}
and the $V$-band magnitudes from the ground-based photometric catalog 
reduced and calibrated following the procedures described in \citet{stetson00,stetson05};\\ 
(3) microturbulent velocities (v$_{t}$) have been obtained by requiring no trend between iron abundances 
and line strengths.

The derived values of $T_\mathrm{eff}$ and v$_{t}$ well agree with those by C13, 
with average differences $\Delta T_\mathrm{eff}$ = $+31\pm8$ K and 
$\Delta v_{t}$ = $-0.03\pm0.01$ km s$^{-1}$.
For log$~g$ there is a systematic difference $\Delta log~g$ = $+0.220\pm0.005$ dex, 
probably due to the different distance modulus and the larger stellar mass adopted by C13.

The abundances of Fe, Na, Mg, and Al have been derived using the classical method of the 
equivalent widths (EW) with
the package GALA \citep{mucciarelli13a}\footnote{http://www.cosmic-lab.eu/gala/gala.php}.
EWs have been measured by means of the DAOSPEC package \citep{stetson08}, iteratively launched with the code
4DAO\citep{mucciarelli13}\footnote{http://www.cosmic-lab.eu/4dao/4dao.php}.
The linelist was built using a synthetic reference spectrum calculated at the UVES resolution
and selecting only transitions predicted to be unblended.
We adopted the atomic data of the last release of the Kurucz/Castelli 
compilation\footnote{http://wwwuser.oats.inaf.it/castelli/linelists.html}
for all species except for FeII lines, which have been taken from \cite{melendez09}.
The adopted model atmospheres have been computed with
the ATLAS9 code\footnote{http://wwwuser.oats.inaf.it/castelli/sources/atlas9codes.html}
adopting a global metallicity of [M/H] = $-1.5$ dex.
The abundances of Na have been corrected for NLTE effects according to \cite{gratton99}
and consistently with the analysis of C13.
For 7 stars the Al lines at 6696-6698$\rm\mathring{A}$ are too weak to be detected
and only upper limits can be obtained.

The abundances of C, N and O have been measured through the spectral synthesis technique, using 
the forbidden oxygen line at 6300$\rm\mathring{A}$, and 
the CH and CN molecular bands at 4300$\rm\mathring{A}$ and 3880$\rm\mathring{A}$, respectively.
To derive the abundance of N we have taken into account the abundance of carbon measured from
the CH band, while for the O abundance we adopted the average C and N abundances thus obtained, 
together with the measured abundance of Ni. This was done  
to take into account the close blending of the O line at 6300$\rm\mathring{A}$ with a Ni transition.
We also checked that the O transition is free from telluric contamination in 19 out of 20 AGB
stars. For the star 1620 the contamination is severe and we did not derive the O abundance.

As reference solar abundances we assumed those of \cite{grevesse98} except for C, N and O,
for which we assumed the values of \cite{caffau11}.

The computation of the final abundance uncertainties adds in quadrature two terms.  
The first is the error arising from spectral features measurements.
For the abundances derived from EWs, 
this term is obtained for each star by dividing the line-to-line dispersion by
the square root of the number of lines used. 
For the elements analyzed with spectral synthesis, the fitting procedure is repeated 
for a sample of 500 synthetic spectra where Poissonian noise has been injected to reproduce 
the noise conditions \citep[see][]{mucciarelli13b}.
The second term is the abundance error arising from atmospheric parameters. 
This has been computed by varying each parameter
by its 1$\sigma$ uncertainty obtained in the analysis. Due to the quality of the spectra we found that
the typical internal uncertainties for $T_\mathrm{eff}$ are lower than $\sim$35K, while for 
log~$g$ and v$_{t}$ we found values lower than 0.1 dex and 0.05 km s$^{-1}$, respectively.

\section{Results}

\subsection{Iron Abundances}
\label{ia}

The derived iron abundance ratios are listed in Table \ref{tab1}
together with the stellar atmospheric parameters.
We obtain average [FeI/H] = $-1.80 \pm 0.01$ dex ($\sigma$ = 0.05 dex) and
[FeII/H] = $-1.58 \pm 0.01$ dex ($\sigma$ = 0.02 dex).
The average [FeII/H] abundance is consistent with the values measured
in RGB stars by \citet{yong03,gratton05,carretta07,carretta09b},
while [FeI/H] is 0.22 dex lower than the metallicity inferred from Fe~II lines.
Such a discrepancy between [FeI/H] and [FeII/H] among AGB stars is too large to be 
explained within internal uncertainties and has been observed previously in other GCs 
\citep{ivans01,lapenna14,lapenna15,mucciarelli15a,mucciarelli15b}.
The same [FeI/H]-[FeII/H] discrepancy remains also if we adopt the atmospheric  
parameters quoted in C13. 
Note that C13 do not measure directly the Fe abundance, but
assume the average RGB [Fe/H] by \citet{carretta07} for all the targets. 
With their atmospheric parameters we derive [FeI/H]=$-$1.77$\pm$0.01 dex ($\sigma$=~0.05 dex) 
and [FeII/H]=$-$1.50$\pm$0.01 dex ($\sigma$=~0.02 dex).
Even if a complete explanation of this effect is still lacking,
this {\sl iron discrepancy} seems to be a general feature of AGB stars in GCs.

\subsection{Light elements}
\label{la}
\begin{figure}[]
\centering
\includegraphics[clip=true,scale=0.70,angle=0]{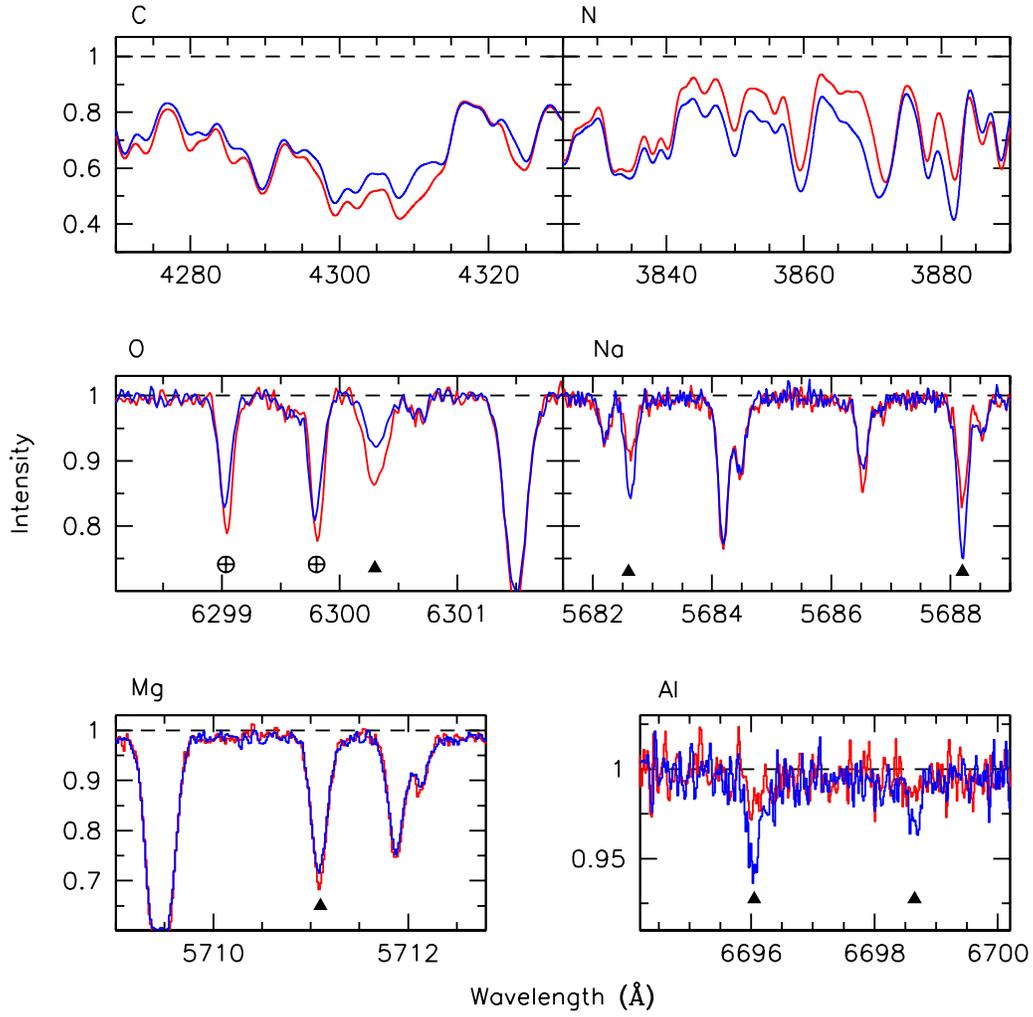}
\caption{Comparison between the spectra of the AGB stars 44 (blue line) and 65 (red line)
in the spectral regions around the atomic and molecular features used in this work 
(and marked with arrow-heads). The black dashed line marks the continuum position. 
In each panel, the black crossed-circles highlight the position of two telluric lines.}
\label{spectra}
\end{figure}

  Significant inhomogeneities in the light element abundances of
  the studied AGB stars are immediately apparent already from the
  visual inspection of the acquired spectra. This can be appreciated
  in Figure \ref{spectra}, where the CH and CN molecular bands,
  and O, Na, Mg and Al lines of star 44 and star 65 (having very similar
  atmospheric parameters; see Table \ref{tab1}) are compared. Apart
  from Mg, notable differences in the line strength are well visible
  for all the other elements. Moreover, the strength of the C and O
  features appears to anticorrelate with the strength of the N and Na
  lines. This clearly shows that the two stars are highly
  inhomogeneous in their light element content.
  
The abundance ratios obtained for the entire AGB sample are
listed in Table \ref{tab1}.  Following the approach discussed in
\cite{ivans01} and \cite{lapenna15}, the abundance ratios have been
computed by adopting [FeI/H] as reference but for O, for which we used
[FeII/H]. This method provides the best agreement between the
abundance ratios in AGB and RGB stars of the same cluster.  However,
because the origin of the FeI-FeII discrepancy is still unclear,
we will discuss the abundances of AGB stars with
respect to both hydrogen and iron, to ensure that our results are
independent of the adopted normalization of the abundance ratios.
  With the only exception of Mg, for which we find values confined
  within a narrow range, the abundances of all the other light
  elements show dispersions well exceeding the internal errors (see
  Table \ref{tab1}). This is true not only for the abundance ratios
  referred to iron, but it also holds for normalizations to
  hydrogen. In particular, the measured sodium abundances span a range
  $\Delta$[Na/Fe]$\simeq\Delta$[Na/H]$\simeq 0.45$, for nitrogen we
  find $\Delta$[N/Fe]$\simeq\Delta$[N/H]$\simeq 0.8$, and for oxygen
  we obtain $\Delta$[O/Fe]$\simeq\Delta$[O/H]$\simeq 0.4$.

\begin{figure*}[ht] 
\begin{center}
\plottwo{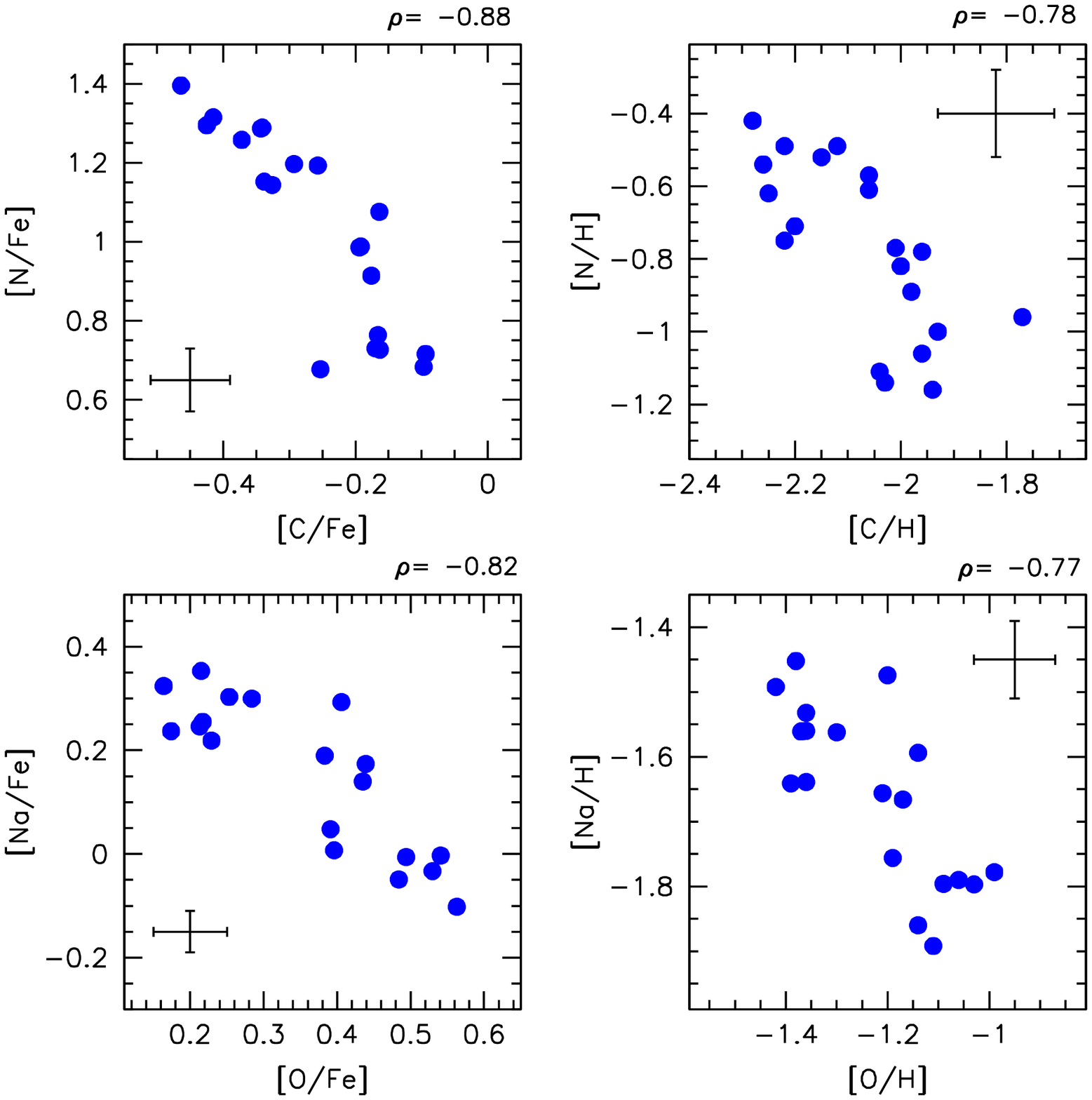}{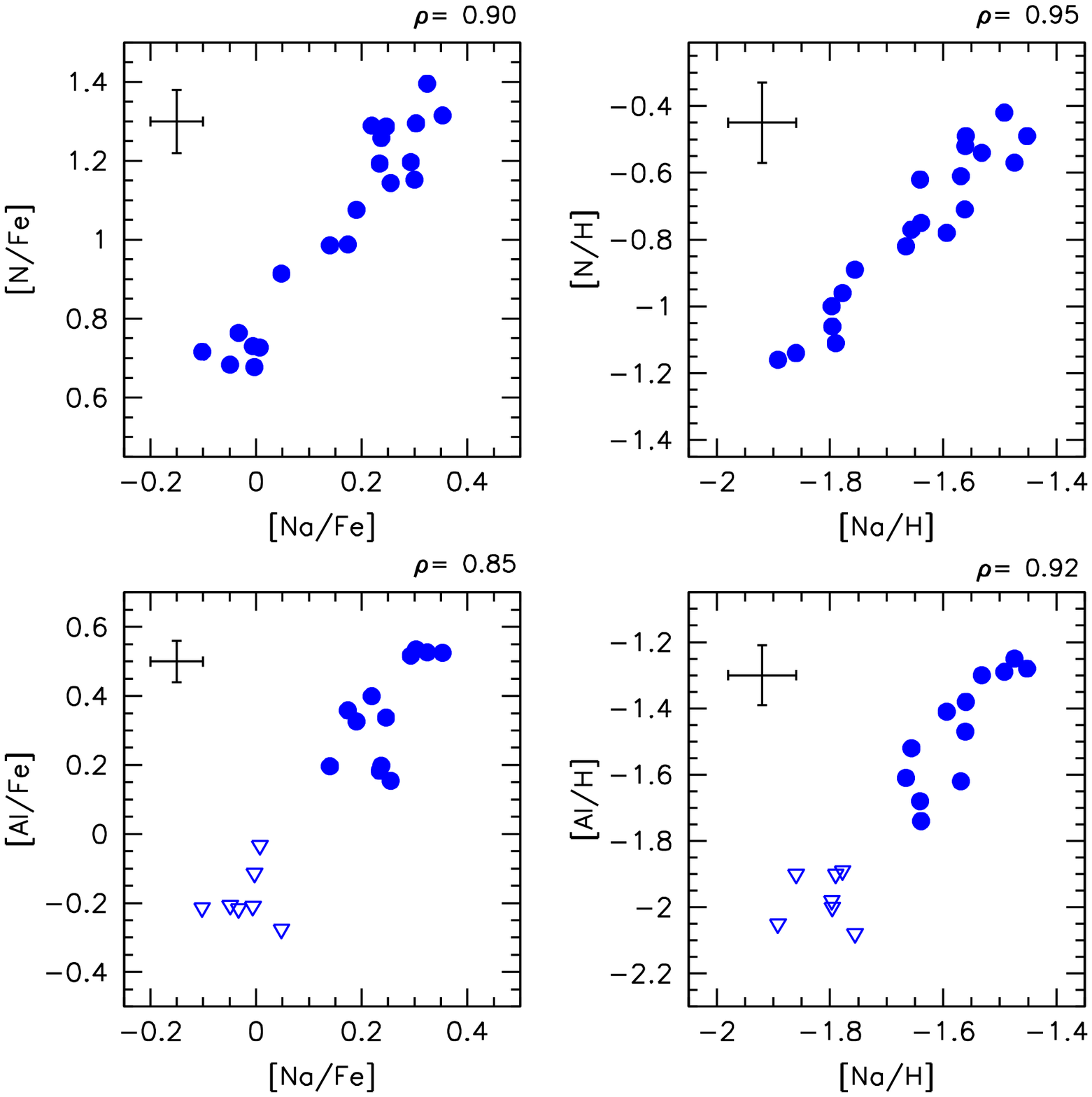}
\end{center}		 
\caption{Light-element abundances measured for the investigated
    AGB stars.  All abundance ratios are shown normalized both to iron
    and to hydrogen.  The typical errorbars of the measured abundances
    and the the Spearman rank coefficients of every correlation are
    marked in each panel.  In the bottom-right panels the empty
    triangles mark the stars for which only upper limits to [Al/Fe]
    have been derived.}
\label{el_corr}
\end{figure*}

  The detected inhomogeneities also appear to be mutually
  correlated. In fact, Figure \ref{el_corr} shows clear C-N and O-Na
  anticorrelations, and N-Na and Na-Al correlations, both if we
  consider the abundance ratios referred to Fe, and if we normalize to
  H. In all cases, the statistical significance, as measured by the
  Spearman rank coefficients $|\rho|$, is very high (values of
  $|\rho|$ larger than 0.74 corresponds to non-correlation
  probabilities lower than $\sim10^{-4}$).  In these diagrams, star 44
  and star 65 (see Figure \ref{spectra}) reside at two opposite ends,
  the former being C and O-rich and N, Na, Al-poor, while star 65
  showing a specular pattern.  The existence of such well-defined
  correlations, by itself, indicates the presence of multiple
  sub-populations along the AGB of NGC 6752. By definition, in fact, a
  sample composed exclusively of FG stars (as suggested by C13) would
  display homogeneous abundances and produce no correlations. Indeed,
  the detected correlations are perfectly in agreement with those
  commonly ascribed to FG and SG sub-populations in GCs (see, e.g.,
  \citealp{carretta09a, carretta09b}).

%%%%%%%%%%%%%%%%%%%%%%%%%%%%%%%%%%%%%%%%%%%%%%%%%%%%%%%%%%%%%%%%%%%%%%%%%%%% DISCUSSION

\section{Discussion}
\begin{figure}[h]
\centering
\includegraphics[angle=0,scale=0.60]{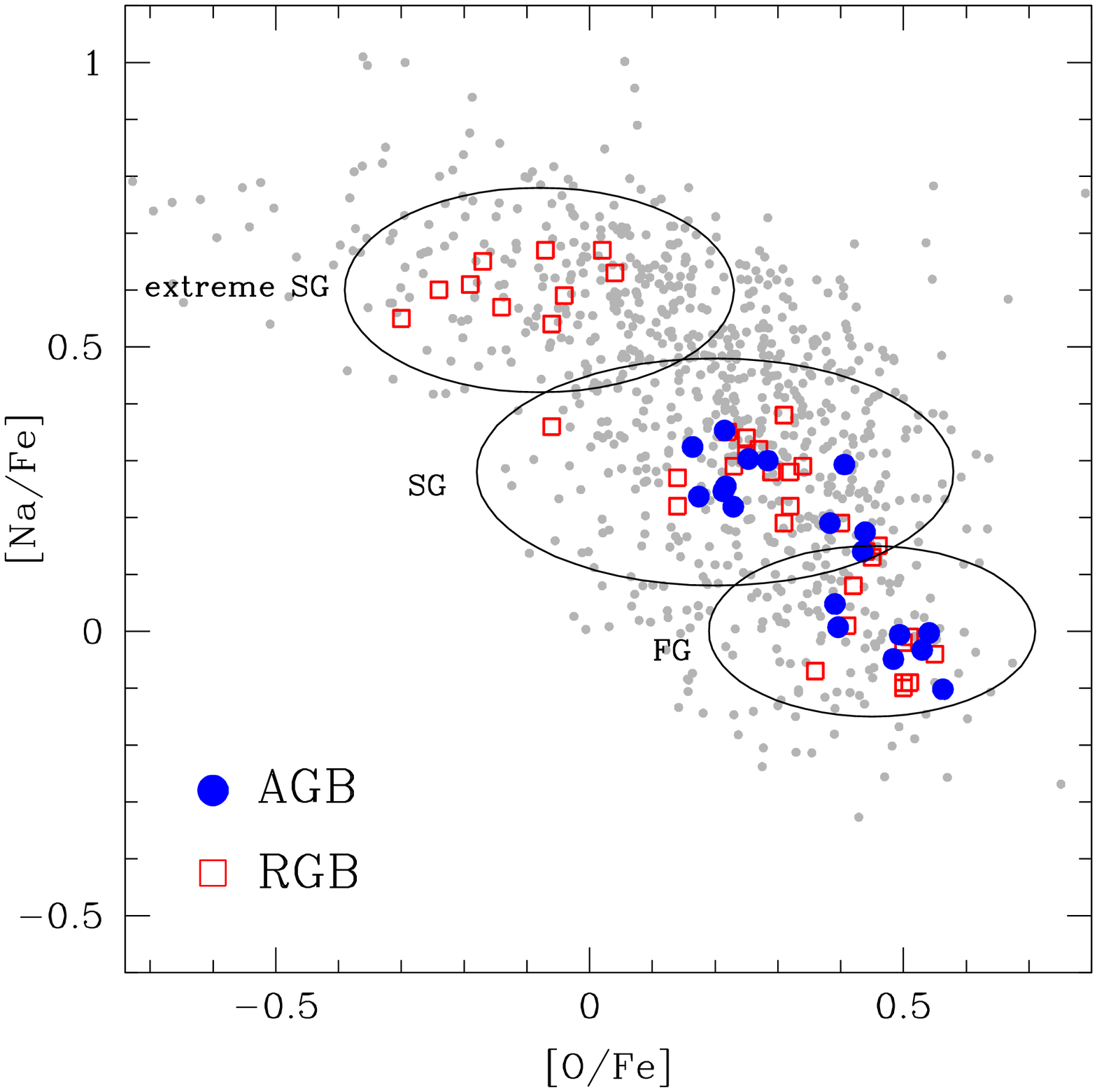}
\caption{Behavior of [Na/Fe] as a function of [O/Fe] for the AGB
  (filled blue circles, this work) and RGB stars \citep[open red
  squares][]{yong03} of NGC~6752. The results obtained for
  RGB stars in other GCs \citep{carretta09a}, rescaled to the solar
  values adopted in this work, are shown as gray dots for
  reference. The regions corresponding to the three populations
  identified by \citet[][ see their Figure 15]{milone13} are encircled.}
\label{nao}
\end{figure}

  Figure \ref{nao} shows the AGB population (solid blue circles) of
  NGC 6752 in the ``standard'' [Na/Fe]-[O/Fe] plane. For reference, we
  also plot the results obtained for the RGB population of NGC 6752
  (empty red squares, from \citealp{yong03}) and several RGB samples
  in 19 GCs (gray dots, from \citealp{carretta09a}). The AGB
  population of NGC 6752 clearly outlines and follows the
  anti-correlation stream defined by the RGB samples, thus confirming
  the existence of SG AGB stars in NGC 6752. To better characterize
  the cluster sub-populations, 
  in Figure \ref{nao} we also plot three
  ellipses corresponding to the values of [Na/Fe] and [O/Fe] that
  \citet{milone13}, on the basis of their photometric study and the
  chemical abundances measured by \citet{yong03}, associate to the FG,
  SG and extreme-SG sub-samples in NGC 6752 (``Populations a, b, and
  c'' in their nomenclature; see Figure 15 in
  \citealp{milone13}). Notably, the abundances here determined for the
  AGB population nicely match the FG and SG loci, thus demonstrating
  that, at odds with the claim by C13, also SG stars do experience the
  AGB phase in NGC 6752.  In particular, based on Table 1 and Figure \ref{nao}, we
  count 13 (out of 20) SG stars, corresponding to $\sim 65\%$ of the
  total AGB sample here investigated.

Following \citet{milone13}, the FG stars in NGC 6752 have standard
chemical mixture ($Y\sim 0.24$), SG stars have moderate enhancement of
[Na/Fe] and He ($Y\sim 0.25$), and a mild depletion of [O/Fe], and
extreme-SG stars have high [Na/Fe] and He ($Y \sim 0.27-0.28$), and
low [O/Fe]. Our results show that the AGB sample is composed of the
first two populations only, while the extreme-SG stars are not
observed. Nicely, the SG/(FG+SG) fraction estimated photometrically by
\citet{milone13} is $\sim 64\%$, in very good agreement with the value
(65\%) found here.  The lack of the extreme-SG stars along the AGB is
exactly what the synthetic HB simulations by \citet{cassisi14}
predict.  These simulations consider three stellar populations with
initial He abundances equal to those quoted by \citet{milone13}, and
they are able to well reproduce the observed value of the $R_2$
parameter and the HB morphology, magnitude and color distribution,
without invoking exceptional mass loss during the HB phase. They
predict that the extreme-SG is the most He-rich population, which
populates the bluer end of the HB and produces {\sl AGB-manqu\'e}
objects.  Also the observed fraction of FG
AGB stars (35\% of the total) is consistent, within the statistical
fluctuations due to the small size of the sample, with the predictions
($\sim 50\%$) of \citet{cassisi14}.

Therefore, the observed fraction of failed AGB stars in NGC 6752 can
be easily explained within the standard stellar evolution framework,
with no need of invoking exceptional mass loss for HB stars hotter
than the Grundahl jump (C13), or extremely high (and inconsistent with
the photometric constraints from the main sequence) initial He
abundances for the SG population \citep{charbonnel13}.

The results presented here provide firm evidence that the AGB
population of NGC 6752 includes both FG and SG stars. While this is in
line with similar results obtained from both spectroscopic
\citep{garcia15} and photometric \citep{milone15a, milone15b,
nardiello15} observations in various GCs, it is in contrast with the findings of C13.

Although a detailed discussion of the origin of the discrepancy with
the result by C13 is beyond the scope of this letter,
a preliminary comparison between the two samples demonstrates that
there is a systematic offset ($\Delta$[Na/Fe] $\sim$ 0.25 dex) between
the values of [Na/Fe] measured here and those measured
by C13\footnote{We also note that C13 adopted  a constant iron abundance for the entire sample.
This can be dangerous in case of AGB stars which do suffer for
the still unclear problem affecting the measure of neutral elemental abundances
\citep[see][]{ivans01, lapenna14, lapenna15, mucciarelli15a, mucciarelli15b}.}.
This is the main reason why while we count at least 11 stars above [Na/Fe] = 0.18 dex
(the threshold adopted by C13), no stars were found by C13.
On the other hand the clear anticorrelations shown in Fig. \ref{el_corr} for the entire set of light elements,
clearly indicated the existence of SG stars along the AGB of NGC6752,
thus demonstrating that the chemical analysis of a single light element
does not allow to draw reliable conclusions about the presence or lack of SG populations.
Indeed, the adoption of the classical scheme based on the analysis of light element (anti)correlations,
appears to be the most appropriate spectroscopic way to detect and distinguish FG and SG stars.

%%%%%%%%%%%%%%%%%%%%%%%%%%%%%%%%%%%%%%%%%%%%%%%%%%%%%%%%%%%%%%%%%%%%%%%%%%%% TABLE

\begin{landscape}
\begin{deluxetable}{lccclclccccc}
\tablecolumns{12}
\tiny
\tablewidth{0pt}
\tablecaption{Atmospheric parameters and abundance ratios of the analyzed AGB stars in NGC~6752.}
\tablehead{\colhead{ID} & \colhead{$T_\mathrm{eff}$} & \colhead{log~$g$} & \colhead{$v_{t}$} &
\colhead{[FeI/H]}  & \colhead{[FeII/H]} & \colhead{[C/Fe]} & \colhead{[N/Fe]} & \colhead{[O/FeII]} & \colhead{[Na/Fe]} & 
\colhead{[Mg/Fe]} & \colhead{[Al/Fe]} \\
& \colhead{(K)} & \colhead{(dex)} & \colhead{(km s$^{-1}$)} & \colhead{(dex)}  & \colhead{(dex)} & \colhead{(dex)} & \colhead{(dex)} & \colhead{(dex)} & \colhead{(dex)} & \colhead{(dex)} & \colhead{(dex)} }
\startdata
   22  &  4554  &  1.18  &  1.85  &  --1.77 $\pm$ 0.01  &    --1.61 $\pm$ 0.01  &     --0.29  $\pm$  0.07  & +1.20  $\pm$  0.12  & +0.41  $\pm$  0.01  &   +0.29  $\pm$  0.04  & +0.37  $\pm$  0.01  &   +0.52  $\pm$  0.02  \\
   25  &  4351  &  0.92  &  1.80  &  --1.79 $\pm$ 0.01  &    --1.60 $\pm$ 0.01  &     --0.25  $\pm$  0.06  & +0.68  $\pm$  0.12  & +0.54  $\pm$  0.06  &  --0.00  $\pm$  0.03  & +0.44  $\pm$  0.01  &  $<$ --0.11  \\
   31  &  4413  &  1.06  &  1.85  &  --1.76 $\pm$ 0.01  &    --1.56 $\pm$ 0.01  &     --0.17  $\pm$  0.07  & +0.76  $\pm$  0.12  & +0.53  $\pm$  0.04  &  --0.03  $\pm$  0.01  & +0.38  $\pm$  0.03  &  $<$ --0.22  \\
   44  &  4585  &  1.31  &  1.65  &  --1.79 $\pm$ 0.01  &    --1.58 $\pm$ 0.01  &     --0.17  $\pm$  0.07  & +0.73  $\pm$  0.12  & +0.49  $\pm$  0.05  &  --0.01  $\pm$  0.03  & +0.40  $\pm$  0.01  &  $<$ --0.21  \\
   52  &  4752  &  1.55  &  1.65  &  --1.81 $\pm$ 0.01  &    --1.61 $\pm$ 0.02  &     --0.19  $\pm$  0.08  & +0.99  $\pm$  0.13  & +0.44  $\pm$  0.04  &   +0.14  $\pm$  0.02  & +0.41  $\pm$  0.01  &   +0.20  $\pm$  0.03  \\
   53  &  4712  &  1.44  &  1.70  &  --1.78 $\pm$ 0.01  &    --1.59 $\pm$ 0.01  &     --0.34  $\pm$  0.08  & +1.29  $\pm$  0.13  & +0.23  $\pm$  0.02  &   +0.22  $\pm$  0.02  & +0.38  $\pm$  0.01  &   +0.40  $\pm$  0.03  \\
   59  &  4724  &  1.50  &  1.65  &  --1.81 $\pm$ 0.01  &    --1.58 $\pm$ 0.01  &     --0.34  $\pm$  0.08  & +1.29  $\pm$  0.13  & +0.21  $\pm$  0.01  &   +0.25  $\pm$  0.01  & +0.38  $\pm$  0.01  &   +0.34  $\pm$  0.05  \\
   60  &  4690  &  1.44  &  1.65  &  --1.68 $\pm$ 0.01  &    --1.55 $\pm$ 0.01  &     --0.09  $\pm$  0.07  & +0.72  $\pm$  0.13  & +0.56  $\pm$  0.05  &  --0.10  $\pm$  0.03  & +0.36  $\pm$  0.02  &  $<$ --0.21  \\
   61  &  4722  &  1.50  &  1.70  &  --1.77 $\pm$ 0.01  &    --1.58 $\pm$ 0.01  &     --0.19  $\pm$  0.08  & +0.99  $\pm$  0.13  & +0.44  $\pm$  0.05  &   +0.17  $\pm$  0.02  & +0.42  $\pm$  0.01  &   +0.36  $\pm$  0.03  \\
   65  &  4622  &  1.31  &  1.80  &  --1.81 $\pm$ 0.01  &    --1.59 $\pm$ 0.01  &     --0.41  $\pm$  0.07  & +1.32  $\pm$  0.13  & +0.21  $\pm$  0.01  &   +0.35  $\pm$  0.03  & +0.40  $\pm$  0.01  &   +0.52  $\pm$  0.03  \\
   75  &  4724  &  1.55  &  1.65  &  --1.84 $\pm$ 0.01  &    --1.59 $\pm$ 0.01  &     --0.10  $\pm$  0.08  & +0.68  $\pm$  0.13  & +0.48  $\pm$  0.03  &  --0.05  $\pm$  0.01  & +0.38  $\pm$  0.02  &  $<$ --0.21  \\
   76  &  4862  &  1.64  &  1.70  &  --1.84 $\pm$ 0.01  &    --1.61 $\pm$ 0.01  &     --0.42  $\pm$  0.08  & +1.30  $\pm$  0.13  & +0.25  $\pm$  0.05  &   +0.30  $\pm$  0.04  & +0.42  $\pm$  0.02  &   +0.54  $\pm$  0.05  \\
   78  &  4877  &  1.65  &  1.75  &  --1.82 $\pm$ 0.01  &    --1.58 $\pm$ 0.01  &     --0.46  $\pm$  0.08  & +1.40  $\pm$  0.13  & +0.16  $\pm$  0.03  &   +0.32  $\pm$  0.03  & +0.38  $\pm$  0.01  &   +0.53  $\pm$  0.04  \\
   80  &  4804  &  1.63  &  1.70  &  --1.80 $\pm$ 0.01  &    --1.58 $\pm$ 0.01  &     --0.18  $\pm$  0.08  & +0.91  $\pm$  0.13  & +0.39  $\pm$  0.02  &   +0.05  $\pm$  0.02  & +0.43  $\pm$  0.01  &  $<$ --0.28  \\
   83  &  4817  &  1.63  &  1.60  &  --1.85 $\pm$ 0.01  &    --1.59 $\pm$ 0.01  &     --0.16  $\pm$  0.08  & +1.08  $\pm$  0.13  & +0.38  $\pm$  0.02  &   +0.19  $\pm$  0.02  & +0.36  $\pm$  0.01  &   +0.33  $\pm$  0.05  \\
   89  &  4798  &  1.63  &  1.65  &  --1.86 $\pm$ 0.01  &    --1.58 $\pm$ 0.01  &     --0.34  $\pm$  0.08  & +1.15  $\pm$  0.13  & +0.28  $\pm$  0.02  &   +0.30  $\pm$  0.02  & +0.38  $\pm$  0.01  &      --      \\
   94  &  4864  &  1.71  &  1.65  &  --1.88 $\pm$ 0.01  &    --1.56 $\pm$ 0.02  &     --0.37  $\pm$  0.08  & +1.26  $\pm$  0.13  & +0.17  $\pm$  0.03  &   +0.24  $\pm$  0.04  & +0.39  $\pm$  0.01  &   +0.20  $\pm$  0.05  \\
   97  &  4884  &  1.75  &  1.70  &  --1.89 $\pm$ 0.01  &    --1.58 $\pm$ 0.01  &     --0.33  $\pm$  0.08  & +1.14  $\pm$  0.13  & +0.22  $\pm$  0.06  &   +0.25  $\pm$  0.01  & +0.40  $\pm$  0.01  &   +0.15  $\pm$  0.06  \\
  104  &  4753  &  1.66  &  1.60  &  --1.87 $\pm$ 0.01  &    --1.54 $\pm$ 0.01  &     --0.16  $\pm$  0.08  & +0.73  $\pm$  0.13  & +0.40  $\pm$  0.07  &   +0.01  $\pm$  0.03  & +0.42  $\pm$  0.01  &  $<$ --0.03  \\
 1620  &  4902  &  1.71  &  1.70  &  --1.80 $\pm$ 0.01  &    --1.56 $\pm$ 0.01  &     --0.26  $\pm$  0.08  & +1.19  $\pm$  0.14  &	   --	       &   +0.23  $\pm$  0.01  & +0.38  $\pm$  0.02  &   +0.18  $\pm$  0.04  \\
\enddata
\tablecomments{Identification number (from C13), $T_\mathrm{eff}$, log~g,
$v_{t}$, and abundance ratios for FeI, FeII, C, N, O, Na, Mg, Al.}
\label{tab1}
\end{deluxetable}
\end{landscape}

%%%%%%%%%%%%%%%%%%%%%%%%%%%%%%%%%%%%%%%%%%%%%%%%%%%%%%%%%%%%%%%%%%%%%%%%%%%%

\acknowledgements
EL acknowledges the financial support from PRIN-INAF 2014.
CL gratefully acknowledges financial support from the European
Research Council (ERC-CoG-646928, Multi-Pop).
DM has been supported by the FIRB 2013 (MIUR grant RBFR13J716).
SC thanks for financial support from PRIN-INAF 2014 (PI: S. Cassisi).
We warmly thank the anonymous referee for his/her useful comments,
which improved the paper.

% ----------------------------- END ---------------------------------- %

{}


\begin{thebibliography}{}

\bibitem[Alonso et al.(1999)]{alonso99}
Alonso, A., Arribas, S., \& Mart{\'{\i}}nez-Roger, C.\ 1999, \aaps, 140, 261

\bibitem[Bastian et al.(2013)]{bastian13}
Bastian, N., Lamers, H. J. G. L. M., de Mink, S. E., 
Longmore, S. E., Goodwin, S. P., \& Gieles, M., 2013, 
\mnras, 436, 2398

\bibitem[Bastian, Cabrera-Ziri \& Salaris(2015)]{bastian15}
Bastian, N., Cabrera-Ziri, \& Salaris, M., 2015, \mnras, 449, 3333

\bibitem[Caffau et al.(2011)]{caffau11}
Caffau, E., Ludwig, H.-G., Steffen, M., Freytag, B., \& Bonifacio, P., 2011, SoPh, 268, 255

\bibitem[Campbell et al.(2012)]{campbell12}
Campbell, S. W., Yong, D., Wylie-de Boer, E. C., Stancliffe, R. J., Lattanzio, J. C., 
Angelous, G. C., Grundahl, F., \& Sneden, C., 2012, ASPC, 458, 205

\bibitem[Campbell et al.(2013)]{campbell13}
Campbell, S.~W., D'Orazi, V., Yong, D., et al.\ 2013, \nat, 498, 198

\bibitem[Carretta et al.(2007)]{carretta07}
Carretta, E., Bragaglia, A., Gratton, R.~G., Lucatello, S., \& Momany, Y.\ 2007, \aap, 464, 927

\bibitem[Carretta et al.(2009a)]{carretta09a}
Carretta, E., Bragaglia, A., Gratton, R.~G., et al.\ 2009, \aap, 505, 117

\bibitem[Carretta et al.(2009b)]{carretta09b}
Carretta, E., Bragaglia, A., Gratton, R., \& Lucatello, S.\ 2009, \aap, 505, 139

\bibitem[Cassisi \& Salaris(2013)]{cs} 
Cassisi, S., \& Salaris, M.\ 2013, ``Old Stellar Populations: How to Study the Fossil Record of Galaxy Formation'', ~Wiley-VCH  

\bibitem[Cassisi et al.(2014)]{cassisi14}
Cassisi, S., Salaris, M., Pietrinferni, A., Vink, J.~S., \& Monelli, M.\ 2014, \aap, 571, AA81

\bibitem[Charbonnel et al.(2013)]{charbonnel13}
Charbonnel, C., Chantereau, W., Decressin, T., Meynet, G., \& Schaerer, D.\ 2013, \aap, 557, LL17

\bibitem[Decressin et al.(2007)]{decressin07}
Decressin, T., Meynet, G., Charbonell, C., Prantzos, N., \& Ekstrom, S., 
2007, A\&A, 464, 1029

\bibitem[Dekker et al.(2000)]{dekker00}
Dekker, H., D'Odorico, S., Kaufer, A., Delabre, B., \& Kotzlowski, H.\ 2000, \procspie, 4008, 534

\bibitem[De Mink et al.(2009)]{demink09}
De Mink, S. E., Pols, O. R., Langer, N., \& Izzard, R. G., 2009, A\&A, 507L, 1

\bibitem[Denissenkov \& Hartwick(2014)]{denis14}
Denissonkov, P. A., \& Hartwick, F. D. A., 2014, MNRAS, 437, 21

\bibitem[Ferraro et al.(1999)]{ferraro99} 
Ferraro, F.~R., Messineo, M., Fusi Pecci, F., et al.\ 1999, \aj, 118, 1738

\bibitem[Garc{\'{\i}}a-Hern{\'a}ndez et al.(2015)]{garcia15}
Garc{\'{\i}}a-Hern{\'a}ndez, D.~A., M{\'e}sz{\'a}ros, S., Monelli, M., et al.\ 2015, \apjl, 815, L4

\bibitem[Gratton et al.(1999)]{gratton99}
Gratton, R.~G., Carretta, E., Eriksson, K., \& Gustafsson, B.\ 1999, \aap, 350, 955

\bibitem[Gratton et al.(2005)]{gratton05}
Gratton, R.~G., Bragaglia, A., Carretta, E., et al.\ 2005, \aap, 440, 901

\bibitem[Gratton et al.(2010a)]{gratton10a}
Gratton, R.~G., D'Orazi, V., Bragaglia, A., Carretta, E., \& Lucatello, S., 2010, \aap, 522, 77

\bibitem[Gratton et al.(2010b)]{gratton10b}
Gratton, R.~G., Carretta, E., Bragaglia, A., Lucatello, S., \& D'Orazi, V.\ 2010, \aap, 517, 81

\bibitem[Gratton, Carretta \& Bragaglia(2012)]{gratton12}
Gratton, R. G., Carretta, E., \& Bragaglia, A., 2012, A\&ARv, 20, 50

\bibitem[Greggio \& Renzini(1990)]{gr}
Greggio, L., \& Renzini, A. 1990, \apj, 364, 35 

\bibitem[Grevesse \& Sauval(1998)]{grevesse98}
Grevesse, N., \& Sauval, A. J., 1998, Space Science Reviews, 85, 161

\bibitem[Johnson et al.(2015)]{johnson15}
Johnson, C.~I., McDonald, I., Pilachowski, C.~A., et al.\ 2015, \aj, 149, 71

\bibitem[Ivans et al.(2001)]{ivans01}
Ivans, I.~I., Kraft, R.~P., Sneden, C., et al.\ 2001, \aj, 122, 1438

\bibitem[Lapenna et al.(2014)]{lapenna14}
Lapenna, E., Mucciarelli, A., Lanzoni, B., et al.\ 2014, \apj, 797, 124

\bibitem[Lapenna et al.(2015)]{lapenna15}
Lapenna, E., Mucciarelli, A., Ferraro, F.~R., et al.\ 2015, \apj, 813, 97

\bibitem[MacLean et al.(2016)]{maclean16}
MacLean, B. T., Campbell, S. W., De Silva, G. M., Lattanzio, J., D'Orazi, V., Simpson, J. D., \& Momany, Y., 
2016, arXiv:1604.05040

\bibitem[Mel{\'e}ndez \& Barbuy(2009)]{melendez09}
Mel{\'e}ndez, J., \& Barbuy, B.\ 2009, \aap, 497, 611 

\bibitem[Milone et al.(2013)]{milone13}
Milone, A. P., et al. 2013, \apj, 767, 120

\bibitem[Milone et al.(2015a)]{milone15a}
Milone, A.~P., Marino, A.~F., Piotto, G., et al.\ 2015, \apj, 808, 51

\bibitem[Milone et al.(2015b)]{milone15b}
Milone, A.~P., Marino, A.~F., Piotto, G., et al.\ 2015, \mnras, 447, 927

\bibitem[Mucciarelli et al.(2013a)]{mucciarelli13a}
Mucciarelli, A., Pancino, E., Lovisi, L., Ferraro, F.~R., \& Lapenna, E.\ 2013, \apj, 766, 78

\bibitem[Mucciarelli et al.(2013b)]{mucciarelli13b}
Mucciarelli, A., Bellazzini, M., Catelan, M., Dalessandro, E., Amigo, P., Correnti, M., 
Cort\'es, C., \& D'Orazi, V., 2013, \mnras, 435, 3667

\bibitem[Mucciarelli(2013)]{mucciarelli13}
Mucciarelli, A.\ 2013, arXiv:1311.1403

\bibitem[Mucciarelli et al.(2015a)]{mucciarelli15a}
Mucciarelli, A., Lapenna, E., Massari, D., Ferraro, F. R., \& Lanzoni, B., \apj, 801, 69

\bibitem[Mucciarelli et al.(2015b)]{mucciarelli15b}
Mucciarelli, A., Lapenna, E., Massari, D., et al.\ 2015, \apj, 809, 128

%\bibitem[Muratov \& Gnedin(2010)]{muratov10} 
%Muratov, A.~L., \& Gnedin, O.~Y.\ 2010, \apj, 718, 1266

\bibitem[Nardiello et al.(2015)]{nardiello15}
Nardiello, D., Piotto, G., Milone, A.~P., et al.\ 2015, \mnras, 451, 312

\bibitem[Norris et al.(1981)]{norris81}
Norris, J., Cottrell, P. L., Freeman, K. C., \& Da Costa, G. S., 1981, ApJ, 244, 205

\bibitem[Renzini et al.(2015)]{renzini15}
Renzini, A., D'Antona, F., Cassisi, S., et al.\ 2015, \mnras, 454, 4197

\bibitem[Stetson(2000)]{stetson00}
Stetson, P. B, 2000, \pasp, 112, 925

\bibitem[Stetson(2005)]{stetson05}
Stetson, P. B, 2005, \pasp, 117, 563

\bibitem[Stetson \& Pancino(2008)]{stetson08}
Stetson, P. B., \& Pancino, E., \pasp, 120, 1332

\bibitem[Ventura \& D'Antona(2005)]{ventura05}
Ventura, P., \& D'Antona, F.\ 2005, \apjl, 635, L149

\bibitem[Vink \& Cassisi(2002)]{vink02}
Vink, J.~S., \& Cassisi, S.\ 2002, \aap, 392, 553

\bibitem[Yong et al.(2003)]{yong03}
Yong, D., Grundahl, F., Lambert, D.~L., Nissen, P.~E., \& Shetrone, M.~D.\ 2003, \aap, 402, 985

\end{thebibliography}
\end{document}